\documentclass[12pt]{article}
\usepackage{amsmath}
\usepackage{graphicx}
\usepackage{enumerate}
\usepackage{natbib}
\usepackage{url} % not crucial - just used below for the URL 

\usepackage[utf8]{inputenc}
\usepackage[T1]{fontenc}
\usepackage{subcaption}

\usepackage{pdflscape}

\usepackage{amsmath,amssymb,amsthm}
\usepackage[colorlinks,linkcolor=blue,citecolor=blue,urlcolor=blue]{hyperref}
\newcommand{\blind}{1}

\newcommand{\bs}{\boldsymbol}

\begin{document}

\def\spacingset#1{\renewcommand{\baselinestretch}%
{#1}\small\normalsize} \spacingset{1}

%%%%%%%%%%%%%%%%%%%%%%%%%%%%%%%%%%%%%%%%%%%%%%%%%%%%%%%%%%%%%%%%%%%%%%%%%%%%%%

\if1\blind
{
  \title{\bf Object oriented data analysis of surface motion time series in peatland landscapes}
 \author{Emily G. Mitchell$^{(1)}$, Ian L. Dryden$^{(2,1)}$, Christopher J. Fallaize$^{(1)}$, \\ Roxane Andersen$^{(3)}$, Andrew V. Bradley$^{(4)}$, David J. Large$^{(5)}$ \\ \& Andrew Sowter$^{(6)}$}
%***
\date{\small $^{(1)}$ School of Mathematical Sciences, University of Nottingham, Nottingham NG7 2RD, UK \\
%***
$^{(2)}$ Department of Mathematics and Statistics, College of Arts, Sciences and Education, Florida International University, Miami, FL 33199, USA \\
%*** 
$^{(3)}$ Environmental Research Institute, University of Highlands and Islands, Castle Street, Thurso, Scotland, KW14 7JD, UK \\
%*** 
$^{(4)}$ Department of Chemical and Environmental Engineering, Faculty of Engineering, Nottingham Geospatial Institute, Innovation Park, Jubilee Campus, Nottingham NG7 2TU, UK \\
%*** 
$^{(5)}$ Department of Chemical and Environmental Engineering, Faculty of Engineering, University of Nottingham, Nottingham, NG7 2RG, UK \\
%*** 
$^{(6)}$ Terra Motion Limited, Ingenuity Centre, Innovation Park, Jubilee Campus, University of Nottingham, Nottingham, NG7 2TU, UK
%***
}

\maketitle
 } \fi

\if0\blind
{
  \bigskip
  \bigskip
  \bigskip
  \begin{center}
    {\LARGE\bf Object oriented data analysis of surface motion time series in peatland landscapes}
\end{center}
  \medskip
} \fi

\vskip 0.5cm 

\begin{abstract}
Peatlands account for 10\% of UK land area, 80\% of which are degraded to some degree, emitting carbon at a similar magnitude to oil refineries or landfill sites. A lack of tools for rapid and reliable assessment of peatland condition has limited monitoring of vast areas of peatland and prevented targeting areas urgently needing action to halt further degradation. Measured using interferometric synthetic aperture radar (InSAR), peatland surface motion is highly indicative of peatland condition, largely driven by the eco-hydrological change in the peatland causing swelling and shrinking of the peat substrate. The computational intensity of recent methods using InSAR time series to capture the annual functional structure of peatland surface motion becomes increasingly challenging as the sample size increases. 
 Instead, we utilize the behavior
of the entire peatland surface motion time series using object oriented data analysis to
assess peatland condition.  In a Gibbs sampling scheme, our cluster analysis based on the functional behavior of the surface motion time series finds features representative of soft/wet peatlands, drier/shrubby peatlands and thin/modified peatlands align with the clusters. The posterior distribution of the assigned peatland types enables the scale of peatland degradation to be assessed, which will guide future cost-effective decisions for peatland restoration.
\end{abstract}

\noindent%
\vfill

\newpage
%\spacingset{1.9} % DON'T change the spacing!

\section{Introduction}
\label{sec:Introduction}

For over 25 years, global peatland restoration has been actively promoted in the hope that the degradation of peatlands can be reversed \citep{verhoeven2014}.
Accounting for one third of Earth’s soil carbon despite only covering 3\% of the land area, peatlands contain up to 95\% water and 5\% organic matter and provide a full spectrum of ecosystem services such as flood regulation \citep{graysonholdenrose2010}, carbon sequestration \citep{pawsonevansallott2012}, and water quality \citep{evanslindsay2010}. Erosion and organic matter loss have a detrimental impact on provision of these services as the peatland system state, or condition, is highly vulnerable to societal pressure, (e.g. grazing, burn-ing, agriculture, forestry, recreation, development and extraction) and susceptible to climate change \citep{isedunn2008,andersen2017,RochefortAndersen2017}.

The UK has 2 Mha of peatlands (10\% land area), mostly as blanket bog, a globally rare type of peatlands. However, up to 80\% of these peatlands are degraded to some degree \citep{bainIUCN2011}. It is estimated that degraded UK peatlands emit 10 Mt C $\text{a}^{-1}$, a similar magnitude to oil refineries or landfill sites \citep{dbeis2019}, placing the UK among the top 20 countries for emissions of carbon from degrading peat \citep{joosten2012}.

To date, condition assessment of all UK peatlands has been impeded by the lack of cost-effective ways to monitor large and often remote areas of peatland.
However recently the use of interferometry synthetic aperture radar (InSAR) signals \citep{sowter2013} to measure peatland surface motion has proved fruitful as a low cost alternative to taking ground measurements \citep{marshall2022}.
Recent literature has found peatland surface motion, a direct consequence of eco-hydrological change in the peatland, to be highly indicative of peatland condition \citep{alshammari2018,alshammari2020,marshall2022,bradley2022}. 
InSAR calculates surface deformation for $80 \times 90 $m$^2$ areas using Sentinel-1 satellites to produce surface motion time series. Surface motion of peatland is mostly driven by the seasonal fluctuation in the water table causing swelling and shrinking of the peat substrate manifesting as an up-down motion (e.g., \cite{howiehebda2017}). In particular, the characteristics (e.g. timing and amplitude of seasonal peaks, overall trend) of these time series are indicative of the peatland condition \citep{bradley2022}. This annual cycle of peatland displacement has been termed ``bog breathing'' \citep{ingram1983hydrology}, and we expect a single peak and a single trough each year in the annual cycle for peatland in good condition \citep{bradley2022}.  
The large volume of data means that current hands-on approaches to data analysis become computationally challenging as study sites increase in area, and so it is of interest to develop a more objective statistical/machine-learning method that would bring more consistency and efficiency to mapping peatland condition.

In this paper we develop statistical methodology for the analysis of InSAR data to assess peatland condition, and in particular we use Object Oriented Data Analysis (OODA) \citep{wang2007,marron2014,marron2021} which provides a framework for analysing data sets of complex objects, such as functions, images and shapes. OODA has strong connections to functional data analysis \citep{ramseysilverman2005}, and indeed the underlying data objects in our study will be a set of functions at distinct spatial locations.

\newpage
\section{The Flow Country and OODA}
\label{sec:Methods}
\subsection{The Flow Country Data}
Covering approximately $4,000$ $\text{km}^2$ of land in the Caithness and Sutherland counties of northern Scotland, the Flow Country is known for being the largest expanse of blanket peatland in Europe \citep{lindsay1988,andersen2018} and the largest carbon store naturally occurring in the UK \citep{chapman2009}. The remote nature of the Flow Country also serves as a refuge for many bird species \citep{lindsay1988}. The current consideration to make the Flow Country a World Heritage site highlights the conservation importance of this area. Compared to other UK peatlands, the Flow Country peatlands remains in good condition overall, however evidence of past interference from land use changes over the last centuries and more recently, including peat cutting, drainage, burn-ing and afforestation still remain. The preservation of the current peatland condition and the reversal of peatland degradation in this region is key for the UK to reduce carbon emissions from peatland.

\begin{figure}[t]
	\begin{center}
	    \includegraphics[scale=0.9]{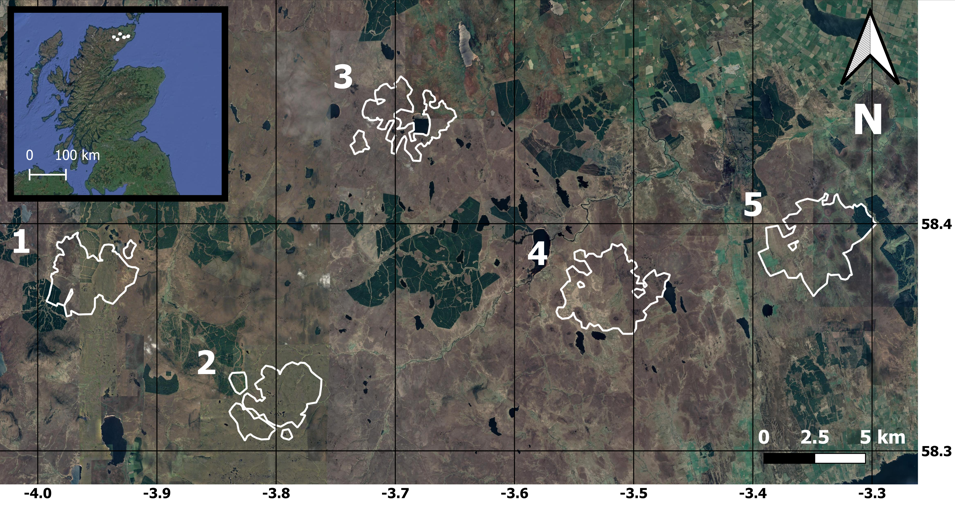}
		\end{center}
		\caption{Location of five peatland sub-sites in the Flow Country, Scotland (inset): Cross Lochs (1), Knockfin (2), Loch Calium (3), Balavreed (4) and Munsary (5). Map grid units in decimal degrees. Base map data: ©2022 Google. Base map imagery: ©2022 TerraMetrics.}
		\label{fig:flowcountrylocs}
\end{figure}

Classifying peatland condition for the Flow Country will indicate which areas are to be targeted for conservation, preservation and restoration according to assigned probabilities. Based on local expert field knowledge, a diverse range of near-natural peatland conditions are captured by five sub-sites in the Flow Country (Figure \ref{fig:flowcountrylocs}), each roughly 10-15 $\text{km}^2$ with unique environmental and management properties \citep{bradley2022}. We assess peatland condition for these five sub-sites using InSAR time series between 12th March 2015 and 1st July 2019. Cross Lochs ($58.39^{\circ}$N, $-3.94^{\circ}$E) at 180 metres above sea level (m.a.s.l), Loch Calium ($58.44^{\circ}$N, $-3.68^{\circ}$E) at 120 m.a.s.l and Balavreed ($58.38^{\circ}$N, $-3.50^{\circ}$E) at 180 m.a.s.l have evidence of low levels of grazing. Cross Lochs and Balavreed consist of flat pool systems whilst Loch Calium has gentle slopes leading down into a central loch. Munsary ($58.39^{\circ}$N, $-3.35^{\circ}$E) at 100 m.a.s.l has been more intensely drained and grazed in the past. Finally, Knockfin ($58.32^{\circ}$N, $-3.80^{\circ}$E) at a much higher altitude (360 m.a.s.l) is an upland plateau with pool systems amongst wind-eroded peat islands, with past grazing and drainage. These peatland areas spanning a wide range of naturally occurring peatlands make them ideal for our analysis based on peatland surface motion.

The coupling between ecohydrological condition and motion of the peat surface \citep{Mahdiyasa2022,marshall2022} make the characteristic time series of surface motion measured by the InSAR technique highly suited to quantifying peatland condition \citep{alshammari2018,bradley2022}. The InSAR signals that measure surface deformation are from the European Space Agency Sentinel-1A and Sentinel-1B satellites and are processed by Terra Motion Limited \citep{sowter2013} to generate a peatland surface motion time series. The surface motion time series are measured at high spatial ($80 \times 90$m$^2$ units) and temporal (every 6-12 days) resolution across the UK. From 12th March 2015, the surface motion time series are every 12 days and an increase in resources with two satellites in operation enabled measurements to be taken every 6 days from 26th September 2016 until 1st July 2019 (Figure \ref{fig:exampleorigtimeseries}). The characteristics of these surface motion time series are indicative of the peatland condition. In total there are $N=9662$ spatial locations in the five regions. 

%% \clearpage

\subsection{Initial OODA questions and answers}
An initial part of OODA often involves asking a set of questions to guide the statistical analysis \citep{marron2021}. In particular we consider the following questions, and provide plausible answers. 

\begin{figure}[t!]
	\centering
	\begin{subfigure}[b]{\textwidth}
   \includegraphics[scale=0.7]{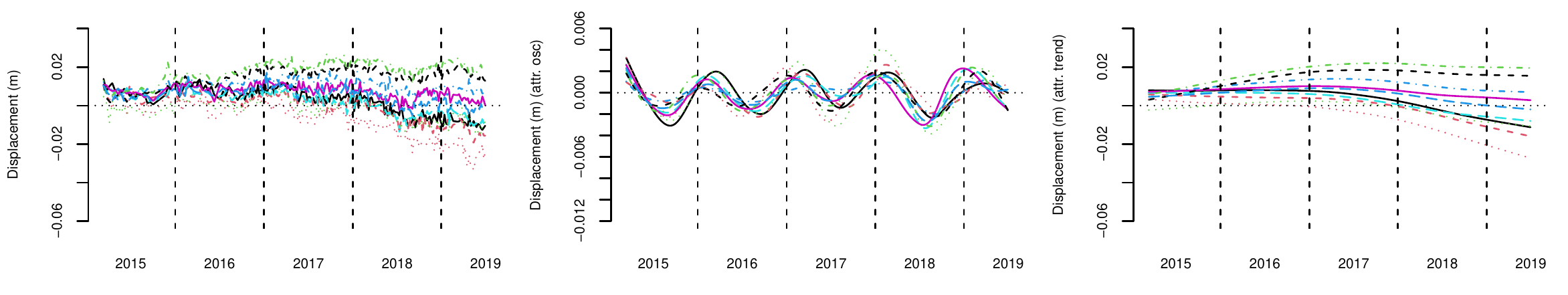}
   \caption{Example locations of soft/wet peatland.}
   \label{fig:Ng1} 
\end{subfigure}
\begin{subfigure}[b]{\textwidth}
   \includegraphics[scale=0.7]{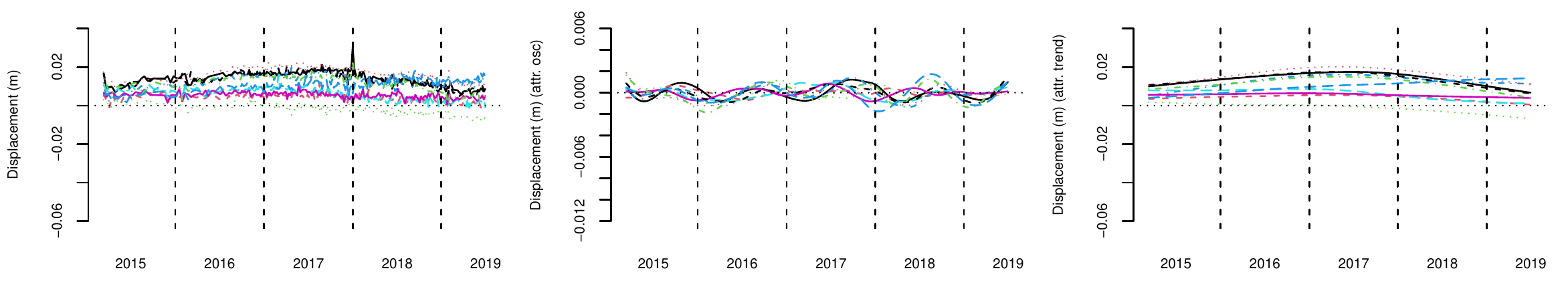}
   \caption{Example locations of drier/shrubby peatland.}
   \label{fig:Ng2}
\end{subfigure}

\begin{subfigure}[b]{\textwidth}
   \includegraphics[scale=0.7]{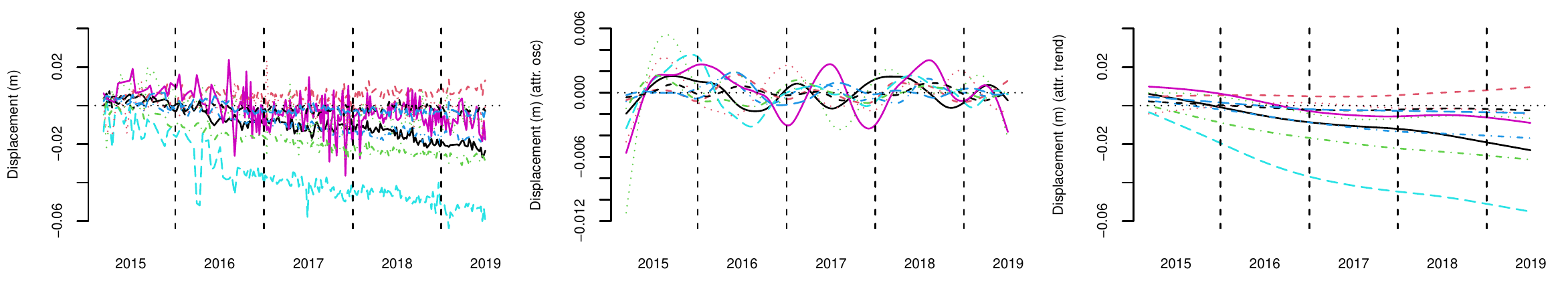}
   \caption{Example locations of thin/modified peatland.}
   \label{fig:Ng3}
\end{subfigure}
	\caption{Left: InSAR time series of peatland surface motion in the Flow Country from 12th March 2015 to 1st July 2019, centre: examples of oscillations, right: examples of trends of peatland surface motion. Time series in each row share oscillatory or trend features, found to be important from past studies. The rows indicate example locations of (1st row) soft/wet peatland, (2nd row) drier/shrubby peatland
and (3rd row) thin/modified peatland. }
	\label{fig:exampleorigtimeseries}
\end{figure}

\begin{enumerate}
\item ``What are the data objects?''

The recorded data are noisy ground displacement time series observed each 6-12 days at a set of spatial locations (where each location is a pixel of area $80 \times 90$m$^2$). However, these are just partial observations of the underlying continuously moving surface. We choose to consider the idealized data objects as functions of continuous time located at discrete locations in space. It is natural to consider the peatland bogs as analogous to ``breathing'' continuously in time, at discrete locations in space.  In order to estimate a smooth function of ground displacement at each location we fit a smoothing cubic spline.

\item ``In what space do the objects lie?''

The idealized object space is a product of function spaces, since at each discrete location we have an underlying function (ground displacement as a function of time). To assess the timing of the annual peaks in the functions we will warp the functions using derivatives with respect to time, and so a suitable object space is the product space of absolutely continuous functions (which are differentiable almost everywhere). 

\item ``What are the important features for practical data analysis?''

From past studies \citep{bradley2022} an overall trend and the timing and amplitude of peak displacements are important for determining peat condition. The overall trend is estimated by fitting a second smoothing cubic spline, with much more smoothing so that the annual cycles are negligible. The de-trended ground  displacement function (termed ``oscillations'') is then obtained by subtracting the trend from the original smoothing spline. In Figure \ref{fig:exampleorigtimeseries} we see some example surface motion time series in the left-hand column; in the middle column are the oscillations for these example locations; and in the right-hand column we see the trends of peatland motion for these examples. The rows of Figure \ref{fig:exampleorigtimeseries} indicate example locations of (1st row) soft and wet peatland, (2nd row) drier, shrubby peatland and (3rd row) thin, modified peatland. For brevity we denote these types of peatland as ``soft/wet'', ``drier/shrubby'' and ``thin/modified'' throughout the paper. It can be seen that the soft/wet peatland has larger amplitude oscillations with earlier timed peaks (in February), the drier/shrubby peatland has smaller amplitude peaks with later timed peaks (in October/November). The thin/modified peaks do not have a consistent structure of oscillations.  The trend is more downward for the thin/modified peatlands whereas it is more flat for the soft/wet and drier/thinner peat here. Hence, the features of amplitude, timing and trend in these examples indicate potential features of interest.   

\item ``What methods will be used?''

The main method that will be used is Bayesian cluster analysis, using a Gaussian model for features based on oscillation amplitude, peak timing and trends. A prior smoothing model for the cluster labels is specified at each location, and the posterior distribution is obtained by Markov chain Monte Carlo simulation. The aim is to produce a map of the area, with estimated class labels indicating similar types of peatland condition, and in addition a measure of the uncertainty in the estimates. 

\end{enumerate}

These OODA questions have been answered over a sustained period by numerous interactions and collaborations with experts on peatland monitoring. The responses to the questions are developed further in the rest of the paper.

%\begin{landscape}
%\end{landscape}

\subsection{Pre-processing} 
 To extract the smooth function for each peatland location, two cubic splines \citep{greensilverman1993,hastietibshiranifriedman2008} are fitted to each time series: one cubic spline for the overall trend and one for the trend and oscillations combined. Furthermore, fitting cubic splines to the data collected every 6-12 days also allows for interpolated daily data to be estimated and outliers to be smoothed over.

Consider the surface motion at each peatland location $j$ for $j=1,\ldots,N$, where the observations at times $t_i \in [a,b], i=1,\ldots,L_j$  are given by $g_{ij}$. The smooth function $\tilde{f}_j$ is found by fitting a natural cubic spline which satisfies
\[\tilde{f}_j = \underset{f_j}{\text{arginf}} \sum^{L_j}_{i=1}\left(g_{ij} - f_j(t_i)\right)^2 + \omega\int^{b}_{a} f_j''(t)^2 dt, \]
where the minimization is over functions $f_j$ which are twice differentiable, and $\omega \ge 0$. There is a trade-off between the first term which is small for a function $f_j$ which closely matches $g_{ij}$ at each time $t_i$ according to their Euclidean distance, and the latter term penalizes for the smoothness of function $f_j$ in the interval  $[a,b]$, where $\omega$ is a smoothing parameter. The cubic smoothing spline enables us to estimate $\tilde{f}_j$ daily. 

For the surface motion time series in the Flow Country sub-sites, we smooth the function between 12th March 2015 ($t_1$) and 1st July 2019 ($t_{202}$), where for all locations $L_j=202$ is the number of time points we have measurements for with either 6 or 12 day spacings.

\begin{figure}[t!]
	\centering
	\begin{subfigure}[b]{0.4\textwidth}
   \includegraphics[scale=0.9]{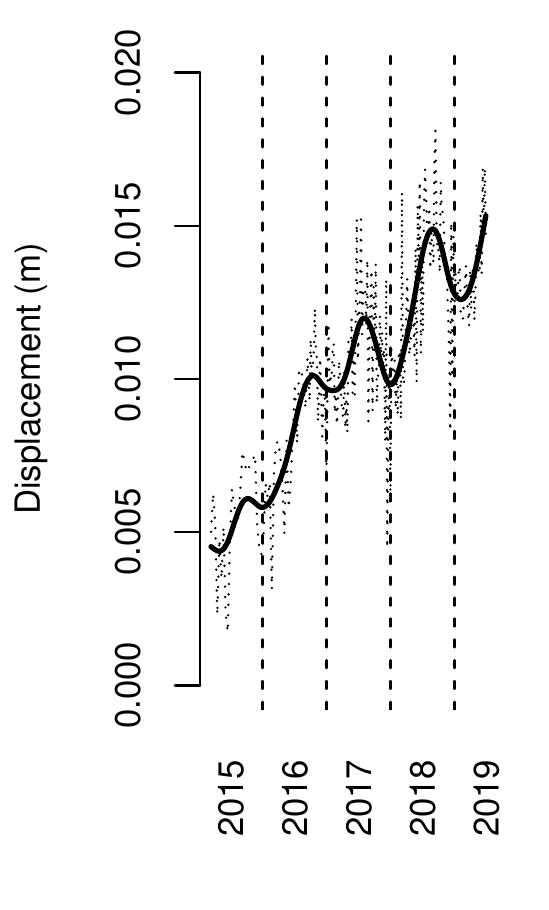}
   \caption{}
   \label{fig:splinedTSFig1} 
\end{subfigure}
\hspace{-1.9cm}
\begin{subfigure}[b]{0.4\textwidth}
   \includegraphics[scale=0.9]{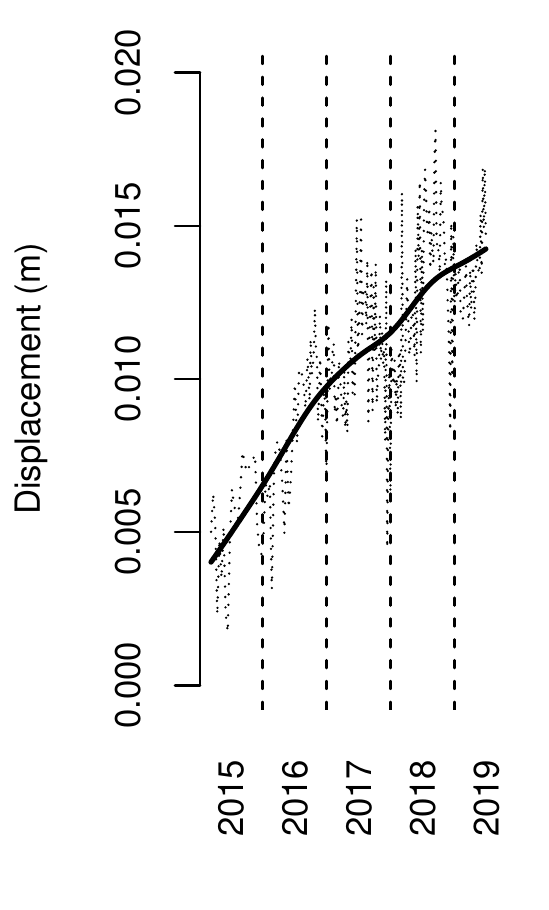}
   \caption{}
   \label{fig:splinedTSFig2}
\end{subfigure}
\hspace{-1.9cm}
\begin{subfigure}[b]{0.4\textwidth}
   \includegraphics[scale=0.9]{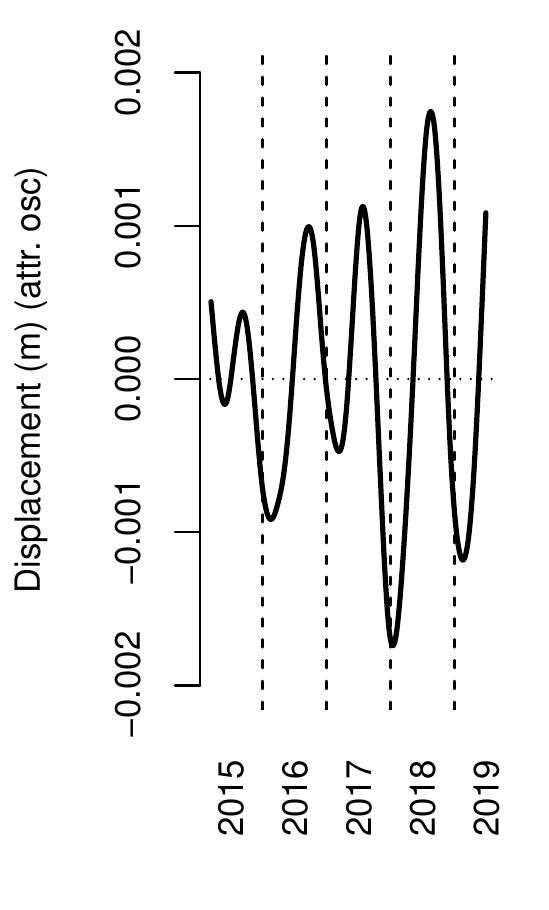}
   \caption{}
   \label{fig:splinedTSFig3}
\end{subfigure}
	\caption{Example of a smoothed peatland surface motion time series. (a) small level of smoothing with ${\tt spar} = 0.7$ to capture trends and oscillations combined, (b) large level of smoothing with ${\tt spar} = 1$ to capture the overall trend, (c) oscillations extracted once the trend has been removed.}
	\label{fig:splineddata}
\end{figure}

The choices of smoothness penalty parameter $\omega$ were made by experimenting with data at many locations. 
An example of the estimated smooth functions for the 
oscillations and trends is given in Figure \ref{fig:splineddata}. 
We use the {\tt smooth.spline} command in R with parameter {\tt spar}, which is a monotonic function of $\omega$. 
When ${\tt spar} = 0.7$ the estimated 
function $\tilde{f}_j$ (Figure \ref{fig:splinedTSFig1}) captures the annual oscillations combined with an overall trend.
When ${\tt spar}=1$  the estimated function $\tilde{f}_j$ estimates the overall trend for each function $j$ (Figure \ref{fig:splinedTSFig2}) without the annual oscillations. The difference between the two smoothed functions gives the de-trended oscillations attributed to bog breathing (Figure \ref{fig:splinedTSFig3}). The smoothed functions of the oscillations and trends will be used throughout, and further examples can be seen in Figure \ref{fig:exampleorigtimeseries}.
Note that we estimate the oscillation plus trend, and trend separately. An alternative approach is to estimate the seasonal and trend components simultaneously allowing for warping, which is the approach considered by \cite{Taietal2017}. 

Since interest lies in the rise and fall of surface motion within the time series itself, the gradient function of the trend is used for the following analysis. The gradient accentuates features interrupting the constant rate accumulation/degradation of peatlands, such as their response to extreme weather events and restoration.

\subsection{Square root velocity functions}

Recall that the OODA approach has led to the data being treated as functions at discrete spatial locations. Functional data analysis \citep{ramseysilverman2005,srivastavaklassen2016} provides a powerful statistical framework to analyse  data as functions of time, where possible warping of the time axis may be needed to register similar features of curves. In our case, functional data analysis allows us to capture the variability in peak timing as well as variability in the amplitude of both the oscillations and the trend gradients. This involves matching each peatland surface motion component to another function by time warping to minimise a distance metric such as the $L_2$ distance. A popular method for warping functions involves first converting to their respective square root velocity function (SRVF) \citep{srivastavawu2011b,srivastavaklassen2011a}. As well as performing well in a wide variety of applications and being straightforward to calculate, the method has appealing theoretical properties. In particular the $L_2$ distance between the SRVFs is equivalent to an elastic metric which is right invariant under simultaneous warping of both functions. 

Without loss of generality, time points $t$ are assumed to lie in $[0,1]$ \citep{srivastavaklassen2011a}. 
For each $j\in\lbrace 1,...,N\rbrace$, suppose the continuous function $f_j$ is defined by $f_j: [0,1] \rightarrow \mathbb{R}$ and the square-root velocity function (SRVF) $q_j: [0,1] \rightarrow \mathbb{R}$ of $f_j$ is given by \begin{equation}
q_j(t) = \frac{\dot{f}_j(t)}{\sqrt{|\dot{f}_j(t)|}}
\label{equ:SRVF}
\end{equation}
when $|\dot{f}_j(t)|\neq 0$ and $0$ otherwise, where $\dot{f}_j$ is the derivative of $f_j$ \citep{srivastavaklassen2011a}. Then the warping function $\gamma_j \in \Gamma$ is chosen by minimising the Euclidean distance between $q_j$ and a template square-root velocity function $q^*$, with a penalty placed on how far the warping function $\gamma_j$ is from the identity:
\begin{equation}
\hat{\gamma}_j=\underset{\gamma_j\in\Gamma}{\text{arginf }}\left|\left|q^* - (q_j \circ \gamma_j) \sqrt{\dot{\gamma}_j}  \right|\right|_2^2 + \lambda \left|\left| 1 - \sqrt{\dot{\gamma}_j}  \right|\right|_2^2,
\label{equ:warpfunc}
\end{equation}
where $\lambda \ge 0$ and the $L_p$ norm is $\| f \|_p = \{ \int_0^1 f(t)^p dt \}^{1/p}$, $p \ge 1$.
Being a popular choice for penalization \citep{srivastavaklassen2016}, the penalty placed on the warping function is based on the squared $L_2$ distance between the first-order derivative of the warping function $\gamma_j$ and $1$, and so this captures the distance to the identity warp (where there is no warping). We denote the $j^{th}$ registered function to the template $q^*$ as $\tilde{q}_j$, where
\begin{equation}
\tilde{q}_j = (q_j\circ \hat{\gamma}_j)\sqrt{\dot{\hat{\gamma}}_j}, \qquad j=1,\ldots,L.
\label{equ:qtilde}
\end{equation}

In our context, the functions to be registered will be oscillation and trend components of the peatland surface motion time series transformed to their SRVF representations. For the oscillations, this will extract the variability in peak timing, captured by $\gamma_j$, from the variability in peak amplitude, captured by $\tilde{q}_j$ from (\ref{equ:qtilde}). 

In practice, the function $f_j$, $j\in \lbrace 1,...,L\rbrace$, is obtained by smoothing the raw data to give estimated values at discrete daily time points $\lbrace 1, ...,1573\rbrace$, with $t=1$ being the 12th March 2015 and $t=1573$ being the 1st July 2019. To improve computational efficiency, each SRVF is sub-sampled at every $20^{th}$ day when computing the warping and distances. 
Examples of the $f_j$ for oscillation and trend functions are seen in Figure \ref{fig:exampleorigtimeseries} displayed at daily time points.

The warping penalty, i.e. the second term of Equation (\ref{equ:warpfunc}), restricts the level of warping permitted to closely match each component of peatland surface motion to their respective reference functions. This allows for some movement in the timings but avoids over-warping to depict the reference functions exactly, which may no longer represent the features in the original function \citep{wusrivastava2011}. Recall that we expect one cycle each year, and over-warping can manifest itself by squeezing together peaks should more than one peak occur annually in the oscillation function. We choose $\lambda=0.1$ to fulfil the trade-off between over-warping to the reference function and negligible warping.

\begin{figure}[t]
	\centering
	\begin{subfigure}[b]{0.5\textwidth}
   \includegraphics[scale=0.9]{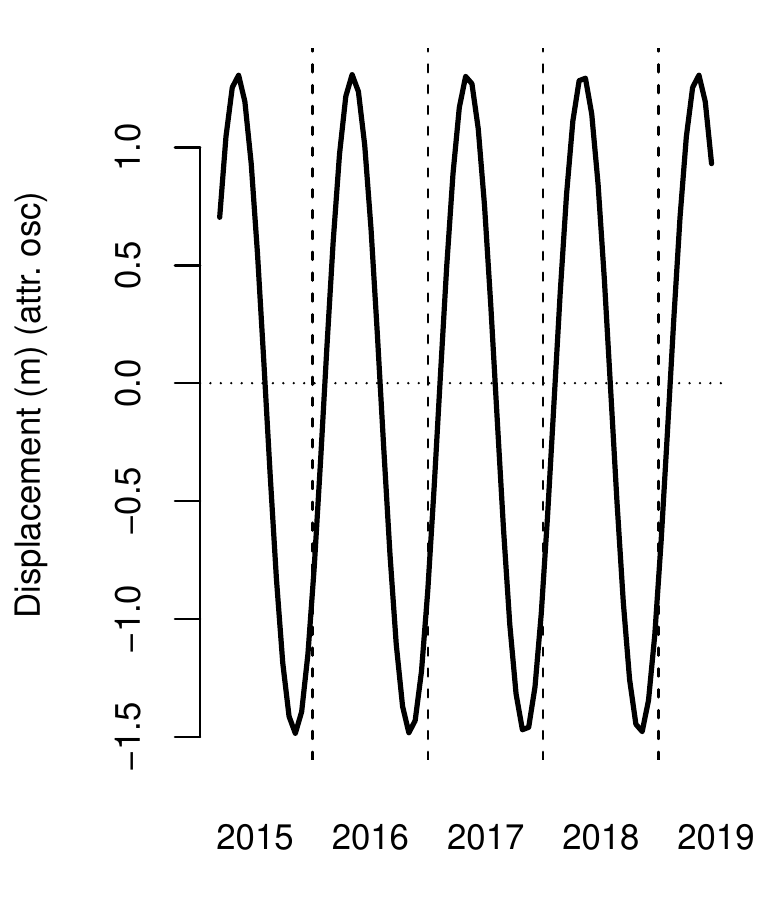}
   \caption{}
   \label{fig:meanshapesFig1} 
\end{subfigure}
\hspace{-1cm}
\begin{subfigure}[b]{0.5\textwidth}
   \includegraphics[scale=0.9]{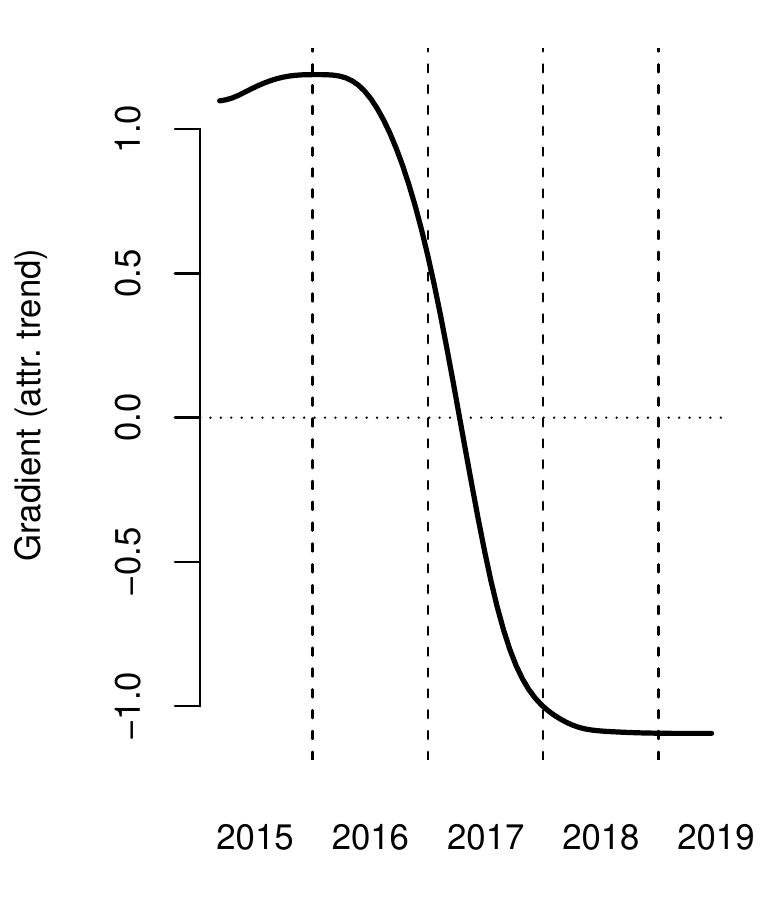}
   \caption{}
   \label{fig:meanshapesFig2}
\end{subfigure}
	\caption{(a) standardized sinusoidal template with peaks at the middle of astronomical spring (roughly 5th May) and troughs at the middle of autumn (roughly 5th November), (b) standardized trend gradient mean.}
	\label{fig:referenceshapesFDA}
\end{figure}

In order to find the warping function associated to each registered function, a choice needs to be made for the reference function $q^*$. For the oscillations, we use a sinusoidal wave with peaks occurring in the middle of astronomical spring every year and troughs appearing in the autumn (Figure \ref{fig:meanshapesFig1}). According to the individual warping functions, registering to peaks in spring will split those peaking in winter, indicative of soft/wet peatlands, and those peaking in summer, indicative of drier/shrubby peatlands responding to hydrological conditions \citep{bradley2022}. Warping functions used to register to the reference function can be found using {\tt pair\_align\_functions} in the R package {\tt fdasrvf} \citep{tuckerwusrivastava2013,fdasrvf2021}. The corresponding registered SRVFs can be found by first converting each function to their SRVF using {\tt f\_to\_srvf}, then warping the SRVF according to the warping function already found using {\tt warp\_q\_gamma}. These functions are also included in the R package {\tt fdasrvf}.

Since little is known about the overall trend for peatlands except there should be very little difference in the timing of the trends within the region, we take the template to be the arithmetic mean of the set of smooth trend gradients, $q^* = \sum^N_{j=1}q_j/N$ (Figure \ref{fig:meanshapesFig2}) and do not consider warping.

\subsection{Distance measures}
Registering each peatland surface motion time series oscillation function to the sinusoidal wave resulted in two functions: the warping function reflecting the difference in timing between the peatland oscillations and the sinusoidal wave and the registered oscillations once the difference in timing has been removed. A similar pair of functions is found for the gradient trends but using the arithmetic mean as the reference function. To classify peatland condition, we look to cluster the peatland locations according to measures computed based on these four functions (three for the oscillations, one for the gradient trends) by comparing each warping function to the identity warping function and comparing each registered function to their respective reference function once standardized in $q$-space. 

\begin{figure}[t!]
	\centering
	\begin{subfigure}[b]{1\textwidth}
   \includegraphics[scale=0.8]{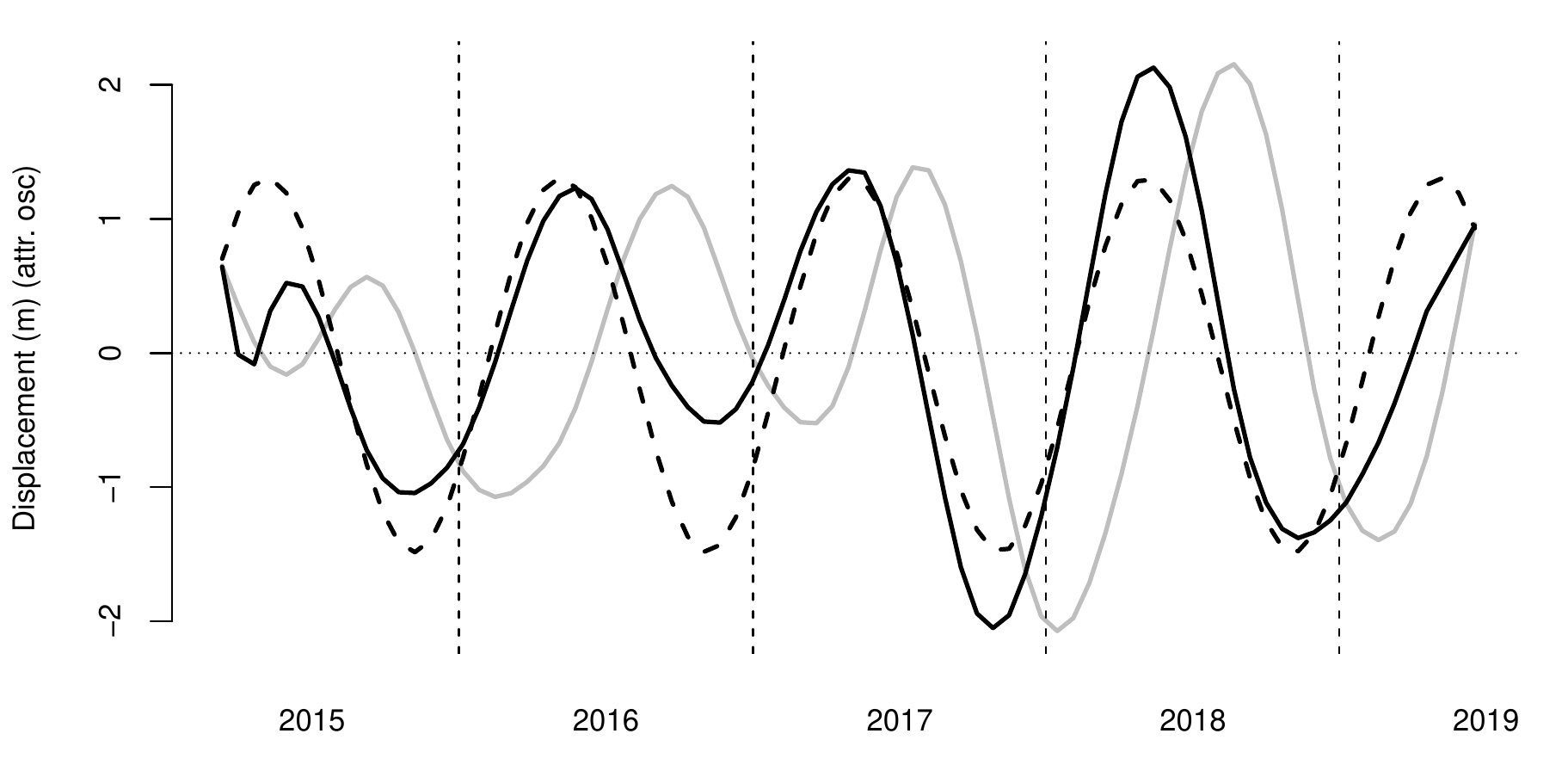}
   \caption{}
   \label{fig:Ng1} 
\end{subfigure}
\begin{subfigure}[b]{1\textwidth}
   \includegraphics[scale=0.8]{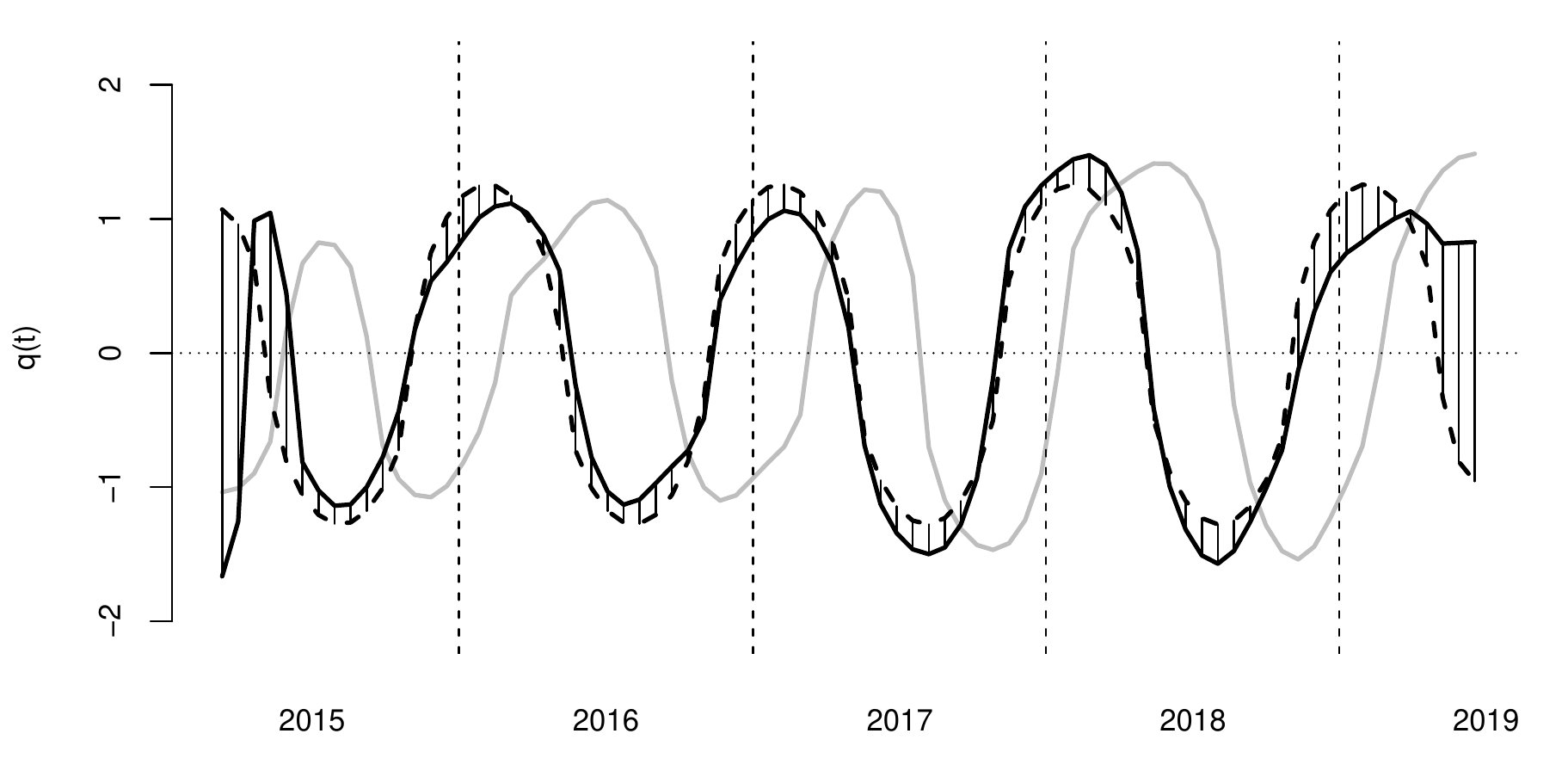}
   \caption{}
   \label{fig:Ng2}
\end{subfigure}
	\caption{(a) the warped oscillations (black solid line) from the original oscillations (grey solid line) to the sinusoidal wave (black dashed line) in the original function space. (b) the warped SRVF (black solid line) from the original SRVF (grey solid line) to the SRVF representation of the sinusoidal wave (black dashed line). The distance measure is calculated between the SRVFs of the sinusoidal wave and the warped oscillations (vertical solid black lines). The $L_1$ norm is taken between the two functions. All functions have been standardized.}
	\label{fig:clustermeasure1}
\end{figure}

The first measure (oscillation amplitude distance) is the $L_1$ distance between the registered oscillation function and the sinusoidal wave once standardized in $q$-space which captures how peat reacts to water storage change \citep{roulet1991,waddington2010}. In order to provide comparable distances over the region each SRVF is standardized to have mean zero and standard deviation 1. An example of a curve, its warp to the template, and the respective standardized SRVF functions is given in Figure \ref{fig:clustermeasure1}. 

\begin{figure}[t!]
	\begin{center}
		\includegraphics[scale=0.8]{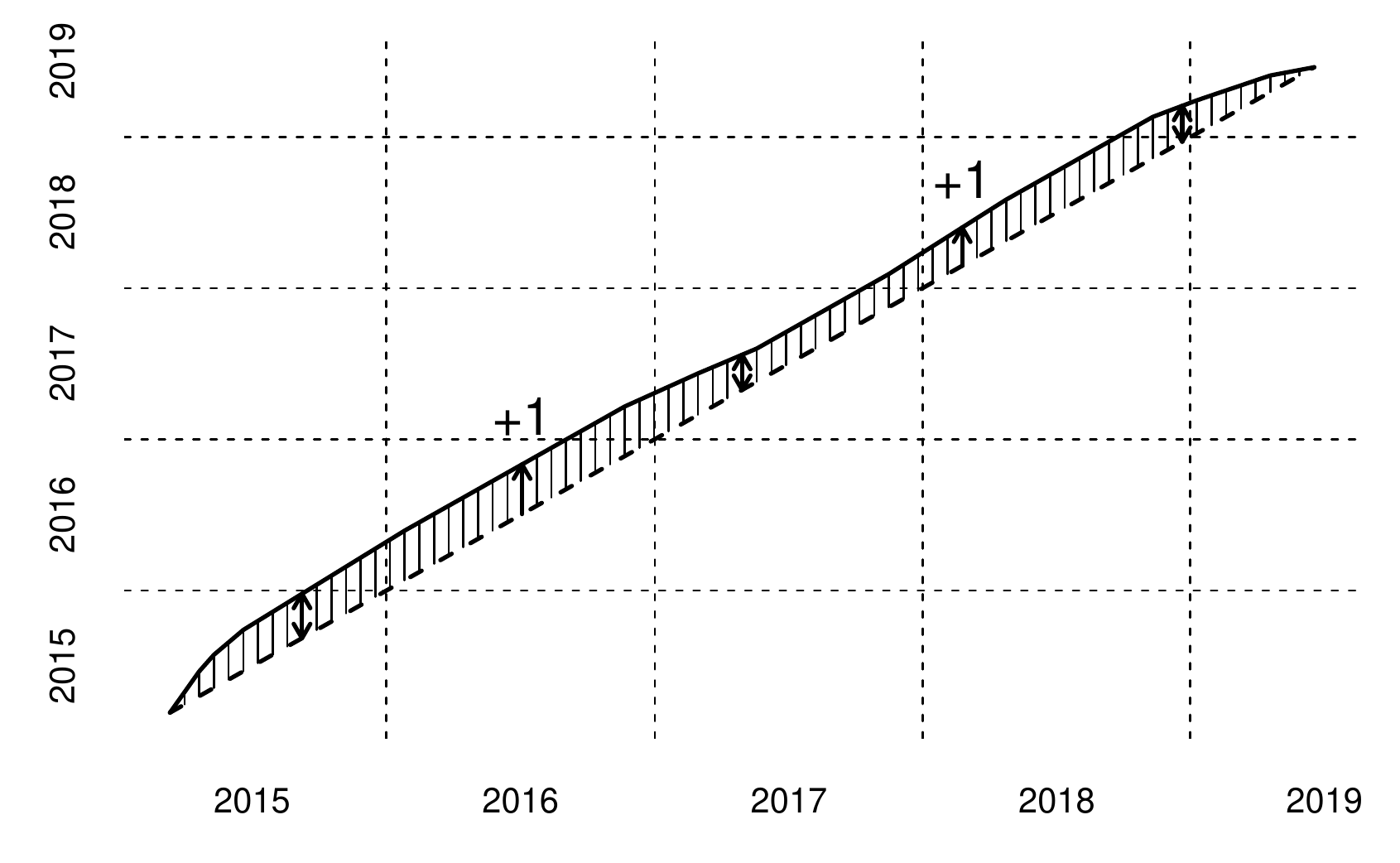}
	\end{center}
	\caption{Warping function (black solid line) to register the time series to the sinusoidal wave, where the identity warping function (black dotted line) represents no warping. The warp distance measure (Table \ref{tbl:1}) calculates the distance between the identity warping function and the warping function to register the time series. This determines the extent to which the timing of the oscillations differ. The warp indicator measure (Table \ref{tbl:1}) indicates whether the registered oscillation function is ahead or behind the timing of the sinusoidal wave. For each time point where the timing is behind, add one. For each time point where the timing is ahead, add zero.}
	\label{fig:clustermeasure3}
\end{figure}

The second measure (oscillation warp distance) and third measure (oscillation warp indicator) 
capture the differences in timing between the oscillation functions and the sinusoidal wave with peaks in the spring,  related to peatland ecohydrology (\cite{alshammari2020}; \cite{tampuu2020}). The warp distance assesses the amount of warping required (Figure \ref{fig:clustermeasure3}) using $L_1$ distance. The warp indicator captures whether the timing of the original peatland surface motion function is ahead, behind or approximately the same as that in the sinusoidal wave. In Figure \ref{fig:clustermeasure3} we see the estimated warp function and the identity warp $\gamma_{id}(t) = t$; the warp distance is the hatched area and the warp indicator is the integral of the distance of all points above identity warp. 

The fourth measure (trend gradient amplitude distance) measures how far the trend gradient SRVF is from the mean trend gradient SRVF, and we use $L_2$ distance so that large departures have more weight. 

 \clearpage

Table \ref{tbl:1} gives details of each equation to compute each measure. The oscillation amplitude distances are calculated using the $L_1$ norm. This will not penalise a peatland surface motion function as a result of our choice of timing of the peaks of the sinusoidal wave as much as if the $L_2$ norm was used. The same applies for the warp distance measure for the oscillations.

\begin{table}[t!]
	\centering
	\begin{tabular}{| l | l  c  c |}
		\hline
		\multicolumn{4}{|l|}{\textbf{Oscillations}} \\
		\hline
%		& & & \\
		1 & Amplitude distance &  $\left|\left|\dfrac{(\tilde{q}_{\text{osc}}-\mu(\tilde{q}_{\text{osc}}))}{\text{sd}(\tilde{q}_{\text{osc}})} - \dfrac{(q^*_{\text{osc}}-\mu(q^*_{\text{osc}}))}{\text{sd}(q^*_{\text{osc}})}\right|\right|_1$ & Figure \ref{fig:clustermeasure1} \\
%		& & & \\
		2 & Warp distance & $\left|\left|\gamma_{\text{id}}-\hat{\gamma}\right|\right|_1$ & Figure \ref{fig:clustermeasure3} \\
%		& & & \\
		3 & Warp indicator & $\left|\left|\text{I}(\gamma_{\text{id}}<\hat{\gamma})\right|\right|_1$ & Figure \ref{fig:clustermeasure3} \\
%		& & & \\
		\hline
		\multicolumn{4}{|l|}{\textbf{Trend gradients}} \\
		\hline
%		& & & \\
		4 & Amplitude distance & $\left|\left|\dfrac{(q_{\text{tre}}-\mu(q_{\text{tre}}))}{\text{sd}(q_{\text{tre}})} - \dfrac{(q^*_{\text{tre}}-\mu(q^*_{\text{tre}}))}{\text{sd}(q^*_{\text{tre}})}\right|\right|_2^2$ & \\
%		& & & \\
		\hline
	\end{tabular}
	\caption{Distance measures based on the registered and warping functions for the oscillations and the registered functions for the trend gradients.}
	\label{tbl:1}
\end{table}

Clustering can now be directly applied to these four measures for the peatland surface motion time series using a common clustering algorithm such as k-means clustering or hierarchical clustering using Ward's method \citep{ward1963}. However, these clusters will not account for the spatial dependency between neighbouring peatland locations; an observed site is more likely to be in the same cluster as another one within its vicinity. To incorporate spatial dependency, we use a Bayesian framework for clustering.

\section{Bayesian cluster and uncertainty analysis}
\subsection{Likelihood and prior distributions} 
Our aim is to carry out cluster analysis so that each peatland location can be assigned to one of several conditions, and also to provide a measure of uncertainty for cluster membership. Since the peatland conditions are expected to be similar in nearby locations we include a spatial prior that encourages nearby locations to be similar.

Let $N$ be the number of peatland locations, $K$ be the number of clusters and $D$ be the number of measures used in our clustering.
We assume the distribution of the features for the $j^{th}$ peatland location, $j\in\lbrace 1,...,N\rbrace$, given membership of cluster $i$, $i \in \left\lbrace 1,... ,K \right\rbrace$, is a $D$-dimensional multivariate Normal distribution with mean vector $\bs{\mu}_i\in \mathbb{R}^D$ and shared cluster variance matrix $\sigma^2\bs{I}_D$, $\sigma^2\in \mathbb{R}^+$, with log density
\begin{equation}
\text{log } p(\bs{X}_j|Z_j=i,\bs{\mu}_i,\sigma^2) = -\frac{D\text{ log}\left(2\pi\sigma^2 \right)}{2} - \frac{1}{2\sigma^2}\left(\bs{X}_j - \bs{\mu}_i\right)^T\left(\bs{X}_j - \bs{\mu}_i\right).
\label{equ:likelihood}
\end{equation}

In this paper, $N=9662$ for the number of peatland locations in the five sub-sites in the Flow Country, $D=4$ for the four measures used in the clustering and the number of clusters is fixed to be $K=3$ to account for the expected types of soft/wet, drier/shrubby and thin/modified peatland observed in the examples of Figure \ref{fig:exampleorigtimeseries}.

To take account of spatial dependency between neighbouring peatlands, we place a Potts prior on the cluster assignment for peatland location $j$ depending on its neighbours $h$ in the neighbourhood $\delta j$ \citep{besag1986}

\begin{equation}
	\text{log } p(Z_j=i) =  \frac{\eta}{|\delta j|} \sum_{h\in\delta j} I(Z_{h}=i) + \text{constant},
	\label{equ:pottsmodel}
\end{equation}
where $\eta \ge 0$ is a fixed parameter determining how much influence the neighbours have in the analysis scaled by $|\delta j|$, the number of neighbours in the neighbourhood $\delta j$. For an observation $h$ to belong in the neighbourhood $\delta j$ of observation $j$, $h$ must be within a specified radius from location $j$. Rescaling $\eta$ to account for how many observations are within the neighbourhood keeps the balance between how much comes from the data and how much comes from the neighbours, such as those on the edge of the regions. In our context, peatlands considered in the neighbourhood of a single peatland location will be in the surrounding circular area of radius $0.00157$ according to their latitude and longitude, on average containing $7$ neighbouring peatlands in the neighbourhood. Furthermore, fixing $\eta = 0.9$ is appropriate for the Flow Country data to make the trade-off between the spatial dependence of neighbouring peatlands and clustering according to the four features.

The log joint distribution of all observations  and cluster labels is
\begin{align*}
& \text{log }p(\bs{X},\bs{Z}|\bs{\mu_1},...,\bs{\mu_K},\sigma^2) = \text{constant} - \frac{DN}{2}\text{ log} \left(\sigma^2\right) \\
& \hspace{2cm} + \sum_{j=1}^N\sum_{i=1}^K I(Z_j=i)\left[ - \frac{1}{2\sigma^2}\left( \bs{X}_j - \bs{\mu}_i\right)^T\left( \bs{X}_j - \bs{\mu}_i\right) + \frac{\eta}{|\delta j|}\sum_{h\in\delta j}I(Z_h = i) \right] .
\end{align*}
Since little is known about the cluster centres $\bs{\mu}_i\in\mathbb{R}^D$, $i\in\lbrace 1,...,K\rbrace$, the prior for each cluster centre $\bs{\mu}_i$ is assumed to be a uniform distribution on a $\mathbb{R}^D$ hypercube  with boundaries defined by the minimum and maximum values of each measure, $\text{log } p(\bs{\mu}_i) = \text{constant}$. This is absorbed into the overall constant term. Finally, for $\sigma^2\in\mathbb{R}^+$ in the shared cluster variance matrix $\sigma^2\bs{I}_D$, to avoid clusters mixing by favouring smaller values for $\sigma^2$ and the added convenience of working with a conjugate prior, $\sigma^2$ is assumed to be distributed according to an inverse Gamma distribution, $\text{log } p(\sigma^2) = -(\alpha+1)\text{ log}\left(\sigma^2\right)-\beta/\sigma^2$, where $\alpha\in\mathbb{R}^+$ and $\beta\in\mathbb{R}^+$ are fixed.
For large $N$ and non-trivial $K>1$, the choices for $\alpha$ and $\beta$ have negligible influence. Commonly used for variance parameters, these are set to be $\alpha = \beta = 0.001$ \citep{spiegelhalter2004,gelman2006}.

After incorporating prior beliefs of the parameters in the mixture model, the log joint posterior distribution is
\begin{align}
&\text{log } p(\bs{X},\bs{Z}|\bs{\mu_1},\ldots,\bs{\mu_K},\sigma^2) + \text{log }p(\bs{\mu_1},...,\bs{\mu_K},\sigma^2) 
= - \frac{DN}{2}\text{log }\sigma^2 - (\alpha + 1) \text{log } \sigma^2 - \frac{\beta}{\sigma^2} \nonumber \\
& + \sum_{j=1}^N\sum_{i=1}^K I(Z_j=i)\left[ - \frac{1}{2\sigma^2}\left( \bs{X}_j - \bs{\mu}_i\right)^T\left( \bs{X}_j - \bs{\mu}_i\right)  + \frac{\eta}{|\delta j|}\sum_{h\in\delta j}I(Z_h = i) \right]  + \text{constant} \label{equ:postdistr} \end{align}
which is then used to sample from the marginal distribution for each parameter in the model: cluster assignments $Z_j$ for $j = 1,... , N$, cluster means $\bs{\mu}_i$ for $i = 1,... , K$ and $\sigma^2$ in the shared cluster variance matrix.

\subsection{Gibbs sampling}
Once initial clusters have been formed using Ward's clustering algorithm \citep{ward1963}, a Gibbs sampler \citep{gemangeman1984} is constructed based on the four measures used for clustering and incorporates the spatial dependency of neighbouring peatlands, enabling samples to be generated from the marginal posterior distributions for the cluster centres and cluster labels.

There are actually three Gibbs steps in each iteration. 
\begin{enumerate} 
\item {\it Cluster assignments}: for the $j^{th}$ observation, 
\begin{align*}
& \text{log }\left( p\left(Z_j | \bs{Z}_{-j}, \bs{X}, \bs{\mu}_1, ..., \bs{\mu}_K, \sigma^2 \right) \right) \\
& \hspace{3cm} = \sum^K_{i=1} I(Z_j=i)\left[ - \frac{1}{2\sigma^2}\left( \bs{X}_j - \bs{\mu}_i\right)^T\left( \bs{X}_j - \bs{\mu}_i\right)  + \frac{\eta}{|\delta j|}\sum_{h\in\delta j}I(Z_h = i) \right] ,\\
& \hspace{4cm} + \text{constant}
\end{align*}
and therefore,
\begin{equation}
Z_j | {\rm rest} \;  \sim \text{Multinomial}(P(Z_j = 1), ..., P(Z_j = i), ..., P(Z_j = K)).
\end{equation}

This is repeated for each observation $j$ in a randomised order resampled each time a cycle of the Gibbs sampler has been run.

\item {\it Cluster means}: Once all cluster labels have been updated, next the cluster means of these are updated for each cluster $i$, $i = 1,...,K$. With the number of observations $n_i$ in cluster $i$,

\[\text{log} \left(p(\bs{\mu}_i| \bs{X},\bs{Z},\sigma^2) \right) = - \frac{n_i}{2\sigma^2} \left(\bs{\mu}_i^T\bs{\mu}_i - \frac{2}{n_i}\bs{\mu}_i^T\sum_{j\in n_i}\bs{X}_j + \frac{1}{n_i}\sum_{j\in n_i}\bs{X}_j^T\bs{X}_j\right) + \text{constant} \]
the log marginal distribution of $\bs{\mu}_i$ is a multivariate Normal distribution
\[\bs{\mu}_i \sim N\left(\frac{1}{n_i}\sum_{j\in n_i}\bs{X}_j,\frac{\sigma^2}{n_i}\bs{I}_D \right).\]

\item {\it Cluster variance}: The final step of the Gibbs sampler is to update the shared cluster covariance matrix $\sigma^2\bs{I}_D$ using the log marginal distribution for $\sigma^2$,

\begin{align*}
& \text{log}\left(p(\sigma^2|\bs{X},\bs{Z},\bs{\mu}_1,...,\bs{\mu}_K)\right) \\
& \hspace{1cm} = -\left(\frac{DN}{2}+\alpha + 1\right)\text{log}\left(\sigma^2\right) - \frac{\beta}{\sigma^2} - \frac{1}{\sigma^2}\sum_{j=1}^N\sum_{i=1}^K\frac{I(Z_j=i)}{2}\left(\bs{X}_j - \bs{\mu}_i \right)^T\left(\bs{X}_j - \bs{\mu}_i \right)\\
& \hspace{6cm} + \text{constant}
\end{align*}
i.e. an inverse Gamma distribution,
\[\sigma^2 \sim \text{IG} \left(\frac{DN}{2}+\alpha ,\beta + \sum_{j=1}^N\sum_{i=1}^K\frac{I(Z_j=i)}{2}\left(\bs{X}_j - \bs{\mu}_i \right)^T\left(\bs{X}_j - \bs{\mu}_i \right) \right).\]
\end{enumerate}

The Gibbs sampler is repeated $T$ times to get parameter samples for $Z_j$ for $j=1,...,N$, $\bs{\mu}_i$ for $i=1,...,K$ and $\sigma^2$ which approximately come from the joint posterior distribution in Equation (\ref{equ:postdistr}), and their respective marginal distributions. For the cluster allocations $Z_j$, $j=1,...,N$, the approximation to the posterior distribution for $Z_j$ allows for probabilities of belonging to each cluster to be found and which cluster each $j$ is likely to be assigned to, i.e. the maximum a posteriori (MAP) estimate. We shall also look at the means of the approximate posterior distributions for the cluster centres to assess what each cluster may represent.

\subsection{Posterior Inference}
We carry out Markov chain Monte Carlo simulation using the multinomial, Gaussian and inverse gamma Gibbs steps, and after many iterations the values of the chain will be (approximate) dependent samples from the posterior distribution. Trace plots are used to identify an appropriate burn-in period, with the remainder of the simulated values used for inference. 

The clustering is estimated by using the maximum a posterior (MAP) estimate, which is the labelling that gives the largest posterior density. In addition we estimate the probability of each location $j$ being in cluster $k$ by observing the proportion of time that that cluster label spends in each cluster after a burn-in period. 

The results will be displayed graphically using a map of the MAP clusters, and individual maps for the probability of the location being in each cluster.

Note that there is a potential for label switching between the clusters, and hence it may be necessary to post-process the MCMC output using an appropriate loss function for the labels.

\section{Results}
We run the MCMC chain for 160,000 iterations and inspect trace plots of the parameters.  For these data the burn-in period seems very short and the chains mix well quickly, but being cautious we take 10,000 as the burn-in period leaving 150,000 observations for the sample. We take every 5th sample to reduce dependency between the samples, which leaves 30,000 samples to base our analysis on. Plots of the sample paths can be found in Appendix A.

\begin{figure}[t!]
	\begin{center}
		\includegraphics[scale=0.8]{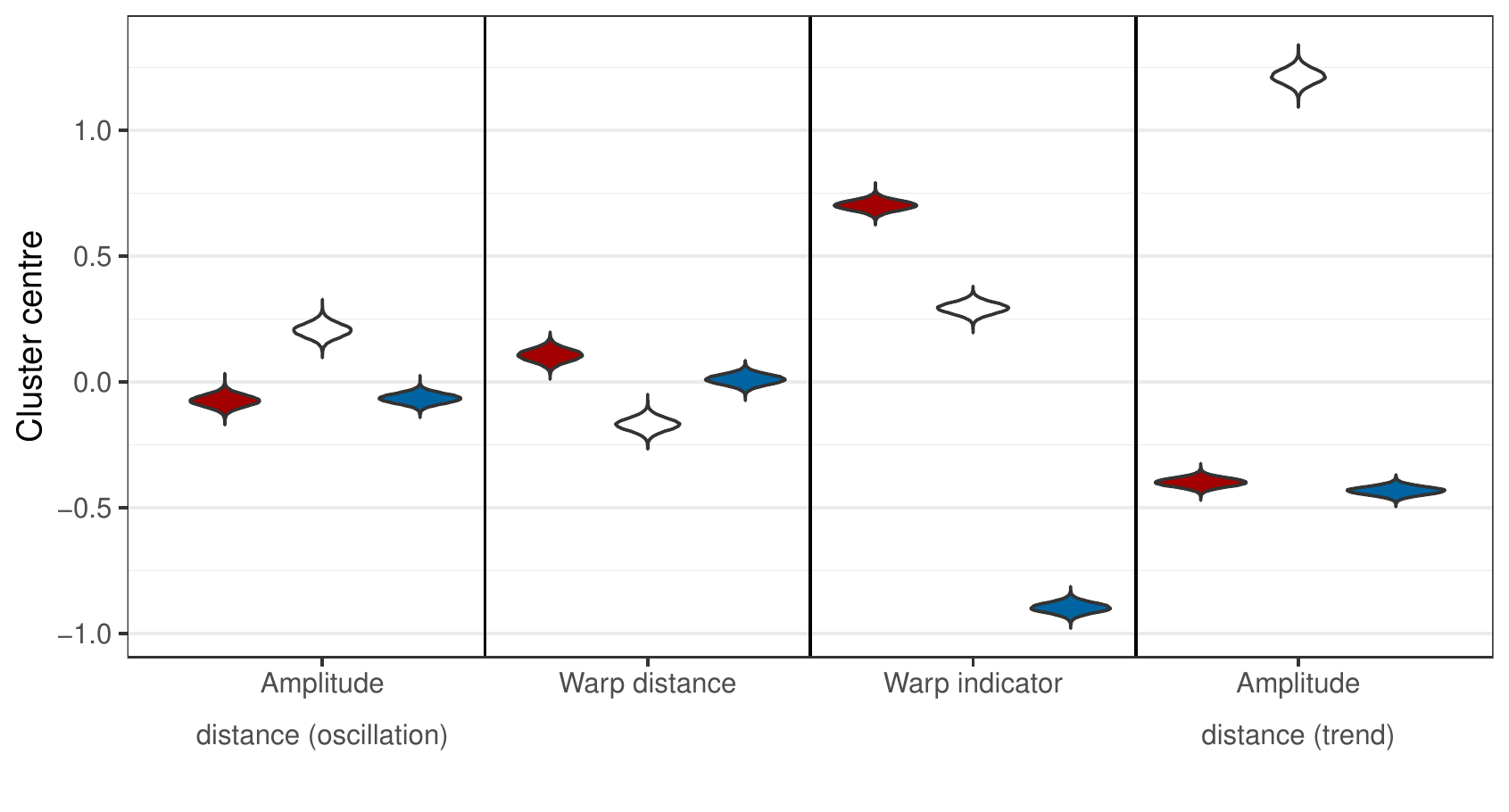}
		\end{center}
		\caption{Violin plots of the posterior means for each of the cluster centres. From the positioning of the posterior means, the blue density represents soft/wet peatland, the red density represents drier/shrubby peatland and the white density represents thin/modified peatland.}
		\label{fig:postmeancentres}
\end{figure}

The posterior means for each cluster mean $\mu_j$ are given in Figure \ref{fig:postmeancentres}. These posterior means help us to identify the type of peatland which is expected in each cluster. 
The posterior mean indicated by the blue density and red density have similar amplitude distances on average for oscillations and trend gradient, but the timing measure by the warp indicator is clearly very different with the blue density being earlier in the year than the sine template peak at 05 May, and the red density being later in the year after 05 May. Hence the blue density cluster is indicative of soft/wet peatlands, whose peaks occur in winter due to hydrological recharge following the growing season \citep{alshammari2020}. The red density cluster, with later peaks, is indicative of drier/shrubby peatlands. The gradient trends for the red density cluster are similar to those in the blue density cluster, both being close to the arithmetic mean and similar long-term change in peatland depth.

The posterior mean of the cluster represented by the white density has a large oscillation amplitude distance suggesting that regardless of how much warping is applied to the oscillations, the oscillations will not closely resemble the sinusoidal wave. The warping indicator is quite close to zero and the warp distance is small suggesting that warping will rarely achieve oscillations close to the sinusoidal wave and so there is little warping to avoid further penalization. There is a very large trend gradient amplitude distance, and so the trend gradient is far from the mean for the region. This cluster is indicative of thin/modified peatlands. Examples for each class can be found in the Figure \ref{fig:exampleorigtimeseries}. 

Samples approximately drawn from the posterior distribution for the cluster labels are also found from the Gibbs sampling scheme. For each of the $N=9662$ peatland locations, the proportion of samples assigned to each cluster will be the probability of being assigned to that peatland type. We also find the maximum a posteriori (MAP) estimate, the peatland type with the highest probability, for each peatland location. For Balavreed, one of the five sub-sites in the Flow Country, the MAP estimates and the corresponding probabilities for each peatland type are plotted in Figure \ref{fig:balavreedMAP}. The plots for the remaining four sub-sites can be found in Appendix B. From extensive expert knowledge and observations that we have of the Flow Country, the identified regions of soft/wet peatlands, drier/shrubby peatlands and thin/modified peatlands, according to the MAP estimates, are very plausible.

\begin{figure}[t!]
\begin{center}
\includegraphics[width=16cm]{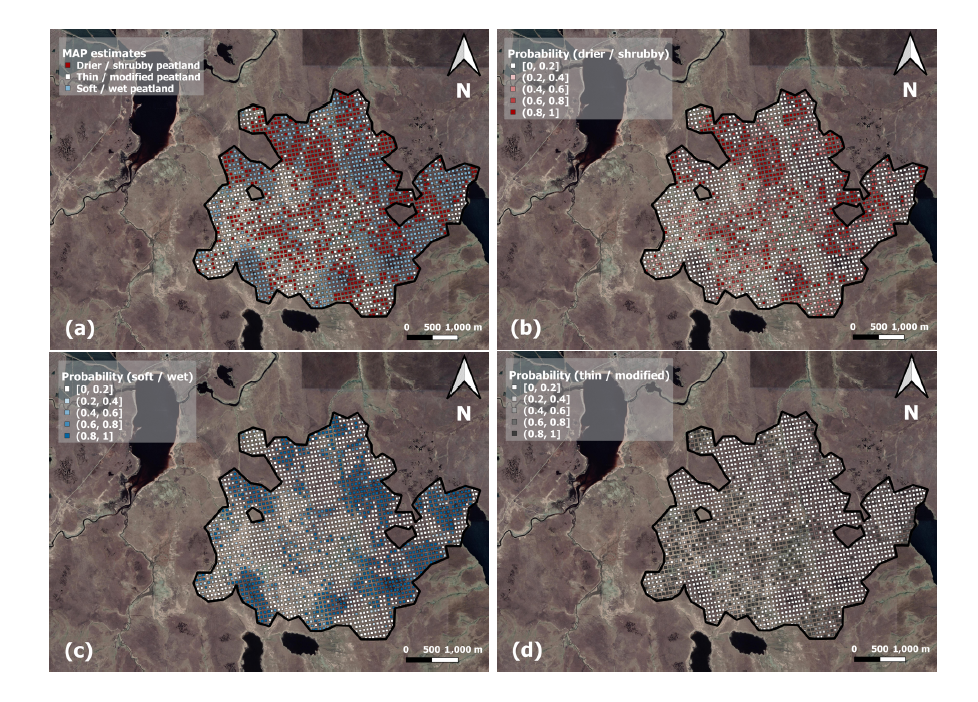}
\end{center}
	\caption{(a) Maximum a posteriori estimate for peatland condition in Balavreed, one of the five sub-sites in the Flow Country. Blue is indicative of soft/wet peatland, red is indicative of drier/shrubby peatland and white is indicative of thin/modified peatland. (b) probability of drier/shrubby peatland, (c) probability of soft/wet peatland, (d) probability of thin/modified peatland (here black is high probability and white low probability). Base map data: ©2022 Google. Base map imagery: ©2022 CNES/Airbus, Getmapping plc, Landsat/Copernicus, Maxar Technologies.}
	\label{fig:balavreedMAP}
\end{figure}

\section{Discussion and conclusions}
\label{sec:Discussion}

Our analysis of peatland surface motion between 12th March 2015 and 1st July 2019 from InSAR data of five sub-sites in Flow Country identified areas of soft/wet, drier/shrubby and thin/modified peatland. Areas identified as thin/modified with a high level of certainty can be earmarked for future restoration and monitored in the meantime. Identified areas of soft/wet and drier/shrubby peatlands must be protected to continue to act as a carbon sink. The areas identified are sensible from ground observations whilst time and expense of conducting field studies are saved. Additionally, this method is far less computationally demanding for larger areas compared to previous methods to assess peatland condition (\cite{bradley2022}).

The first part of the method involved constructing measures based on distances between the registered oscillations and an oscillatory template and the trend gradients and a trend template. To register the oscillations, a reference function was required to base our analysis on. From expert knowledge of peatlands, we used a sinusoidal template with peaks in the middle of astronomical spring and troughs in the middle of autumn which is known to be midway between distinct types of peat: shrubby marginal peats peak in late summer early autumn and soft wet sphagnum peaks in mid winter. It would be unusual for the peatland surface motion to be exactly represented by this wave once warped, nevertheless we would expect something close to this for the warped versions of the soft/wet and drier/shrubby peatlands due to the hydrological recharge of the shrubs in the dry peatlands during the summer growing season and the soft/wet peatlands during the winter. 

To find the peatland oscillation motion attributed to the peat substrate and not the motion from the underlying hydrology, it would be useful to gather precipitation data for the area and incorporate this into our preconception of the local oscillations for specific years. 

An alternative approach, which would avoid selecting the reference function, is to take the reference function as the Karcher mean \citep{srivastavawu2011b} of the set of oscillation SRVF functions. However some concerns include the influence non-oscillatory regions such as the thin/modified peat, and also the effect of outliers.

We restricted the level of warping for the oscillations using smoothness penalty $\lambda = 0.1$ to allow peak timing to be shifted, at most, approximately half a year. It would be unreasonable to suggest the timing between two peatland surface motion time series differed by over half a year since we expect the oscillatory cycle to repeat annually. Further work could entail switching to a Bayesian approach for functional data analysis \citep{chengdryden2016} and putting a prior such as half-Cauchy on $1/\lambda$ to favour larger levels of $\lambda$.

In addition to the oscillations, our analysis is also based on the trend gradients, where we chose to not include any warping. Little is known about the trends in the region except that there should be very little difference in the timing of the trends. Hence there is very little difference between taking the arithmetic mean or the Karcher mean as the reference function, but finding the Karcher mean is far more computationally demanding.

One important question is whether the OODA part of our method, based on the five sub-sites in the Flow Country, can be extended to other regions. These sites were selected because they contained varied blanket peatland conditions in a near-natural state. For other regions, the underlying analysis of the oscillations based on a fixed sinusoidal template can be transferred. The overall trend was based on the arithmetic mean of the trend gradients of the surface motion time series in the five sub-sites. The speed at which the overall trend declines within these regions increases until late 2017, where it continues to rapidly decrease at the same pace. This could be a result of extreme weather events during 2017 and 2018 from which the peatland has yet to recover \citep{fenner2011,stirling2020,undorf2020}. If the trend gradients were included in the clustering, a different region, which may not have exhibited such weather events, with different trend gradients would be classed as degraded when compared to the Flow Country. Topography, climatology and geography of a landscape can all influence peatland surface motion. A potentially fruitful area of research work would be to study different sites with varied peatland conditions, climate and topography and compare their means.

The second part involved clustering according to the distance measures based on the oscillations and the trend gradients. We fixed the number of clusters in this paper to be $K=3$, to reflect our knowledge of the peatlands largely consisting of soft/wet peatlands, drier/shrubby peatlands and thin/modified peatlands. If we were to move to a different region, those which have a significantly different trend gradients may be classed as thin/modified, although they may still contain the oscillatory behavior found in soft/wet or drier/shrubby peatlands. A potentially fruitful area of work would be to investigate the number of clusters, either by penalizing for the number of clusters included or creating a tree structure, by first classifying according to trend gradients followed by classifying according to oscillations. Penalization can be incorporated as a Poisson or Gamma prior placed on the number of clusters in the MCMC sampler when smoothing to account for neighbours.

Another interesting question, which is beyond the scope of the present paper, is to assess the peatlands' response to restoration by examining the behavior of the function before restoration and the behavior afterwards. Testing the success of restoration would instead require a sliding window approach to find if and when the functional behavior changes. In addition to restoration success, a sliding window would allow peatland condition to be assessed through time. If we continue to use a sinusoidal wave as the template for the oscillations and cluster according to the measures inside the sliding window, the changepoint could be identified as the point where the peatland type probabilities significantly change. This would not only enable reflection on the success of past restoration, it would also indicate best restoration practices for varied types of peatland environments, to support future action.

We have provided a new method for the analysis of environmental time series which would also be appropriate in many more general applications, such as monitoring ocean temperatures, ENSO, phenology, and gas emissions, where seasonal cycles and trends are present. Using the same Object Oriented Data Analysis approach to building appropriate methodology is anticipated to be a valuable approach in such applications.

\if1\blind
{\section*{Acknowledgements}

%\section*{Funding}

%*** 
This work has been funded by the Natural Environment Research Council (NERC), grant number NE/T010118/1, as part of the project ``Developing a statistical methodology for the assessment and management of peatland (StAMP)'' in the Landscape Decisions Programme. Roxane Andersen is also supported by a Leverhulme Leadership Award (RL-2019-002).
}\fi

%\section*{Data and code}
%The raw InSAR surface motion time series for the five sub-sites studied in this paper will be provided.
%The R code to produce the analyses of the data discussed in this article will also be provided.

\bibliographystyle{apalike}

\bibliography{reflist}

\renewcommand{\thesection}{\Alph{section}}

\newpage

\section*{Appendix}

Appendix Section A gives some trace plots from the Gibbs sampler output used to approximate the joint and respective marginal posterior distributions. Appendix Section B presents plots of the landscape for the maximum a posteriori (MAP) estimates and probability plots for the sub-sites not given in the manuscript.

\section*{A: Gibbs sampling chains}

\begin{figure}[h]
\begin{center}
\includegraphics[width=16cm]{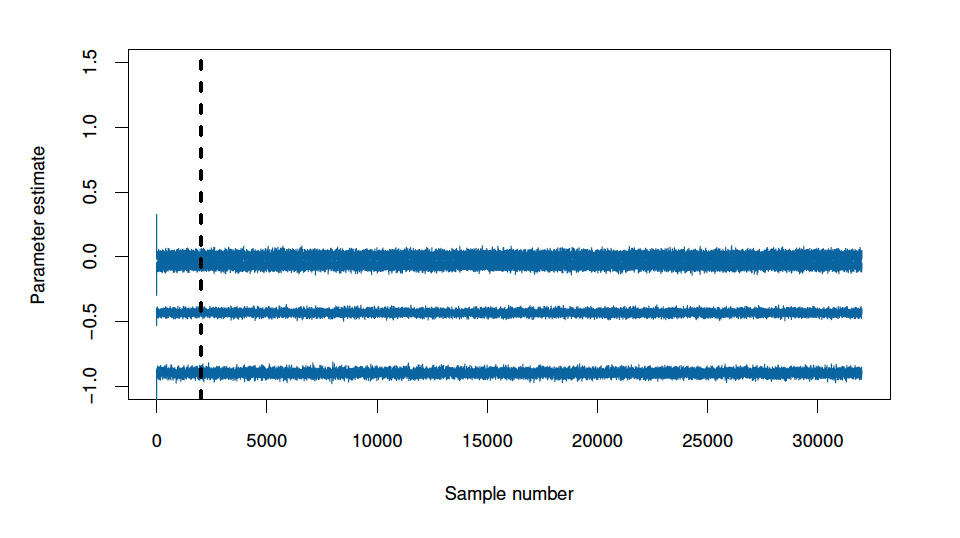}
\end{center}
		\caption{Thinned Gibbs samples for the cluster mean of the cluster aligned with soft/wet peatland, clustered according to the oscillation amplitude distance (line with mean $-0.0656$), warp distance (line with mean $0.0096$), warp indicator (line with mean $-0.8973$) and the trend gradient amplitude distance (line with mean $-0.4320$). Left of the vertical dashed line at $2000$ indicates the burnin period after thinning.}
\end{figure}

\begin{figure}[h]
\begin{center}
\includegraphics[width=16cm]{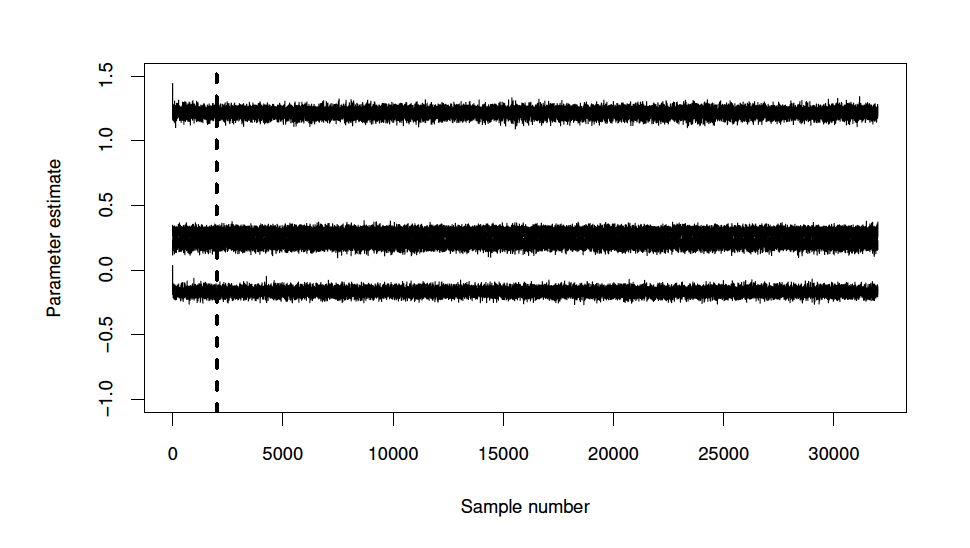}
\end{center}
		\caption{Thinned Gibbs samples for the cluster mean of the cluster aligned with thin/modified peatland, clustered according to the oscillation amplitude distance (line with mean $0.2050$), warp distance (line with mean $-0.1685$), warp indicator (line with mean $0.2933$) and the trend gradient amplitude distance (line with mean $1.2141$). Left of the vertical dashed line at $2000$ indicates the burnin period after thinning.}
\end{figure}

\begin{figure}[h]
\begin{center}
\includegraphics[width=16cm]{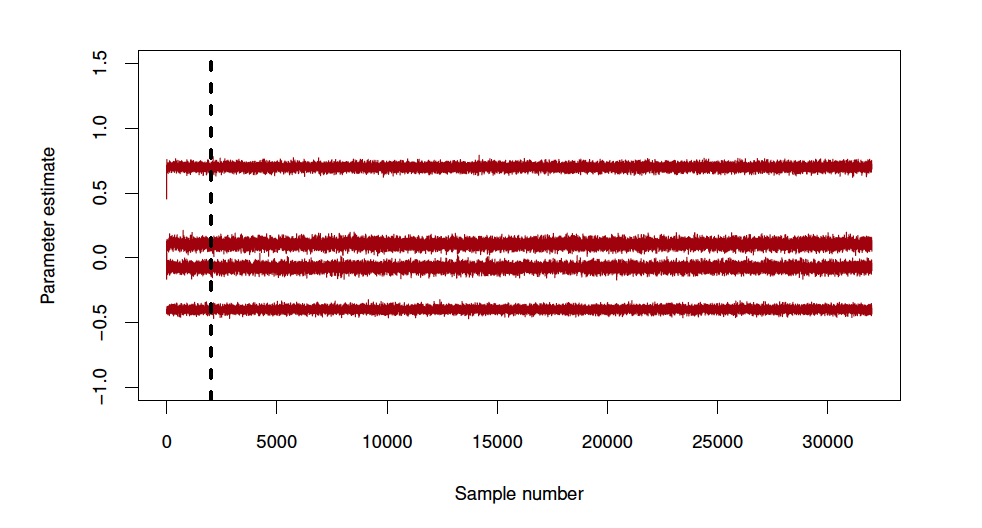}
\end{center}
		\caption{Thinned Gibbs samples for the cluster mean of the cluster aligned with drier/shrubby peatland, clustered according to the oscillation amplitude distance (line with mean $-0.0751$), warp distance (line with mean $0.1064$), warp indicator (line with mean $0.7014$) and the trend gradient amplitude distance (line with mean $-0.3990$). Left of the vertical dashed line at $2000$ indicates the burnin period after thinning.}
\end{figure}

\begin{figure}[h]
\begin{center}
\includegraphics[width=16cm]{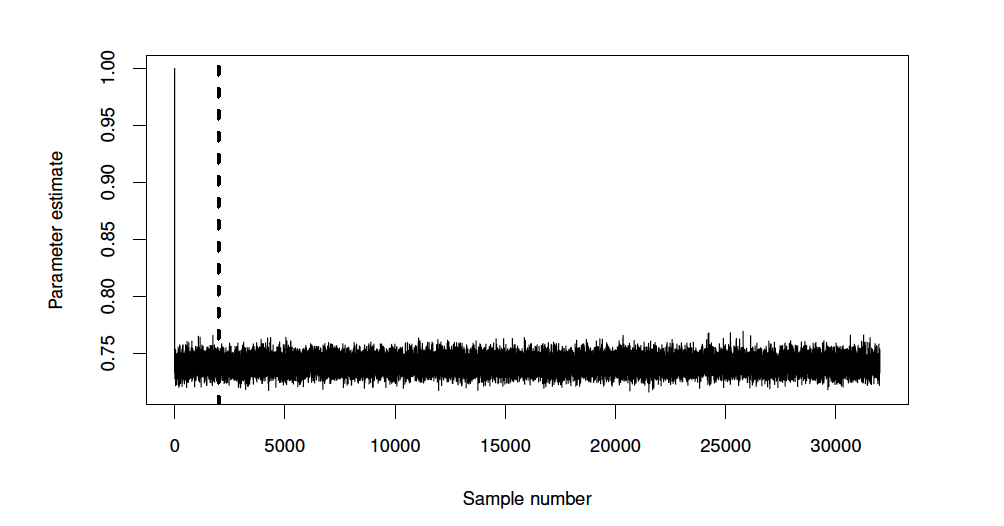}
\end{center}
		\caption{Thinned Gibbs samples for $\sigma^2$ in the shared cluster variance matrix $\sigma^2I_4$. Left of the vertical dashed line at $2000$ indicates the burnin period after thinning.}
\end{figure}

\clearpage

\section*{B: MAP and probability plots for remaining sub-sites}

\begin{figure}[h]
\begin{center}
\includegraphics[width=16cm]{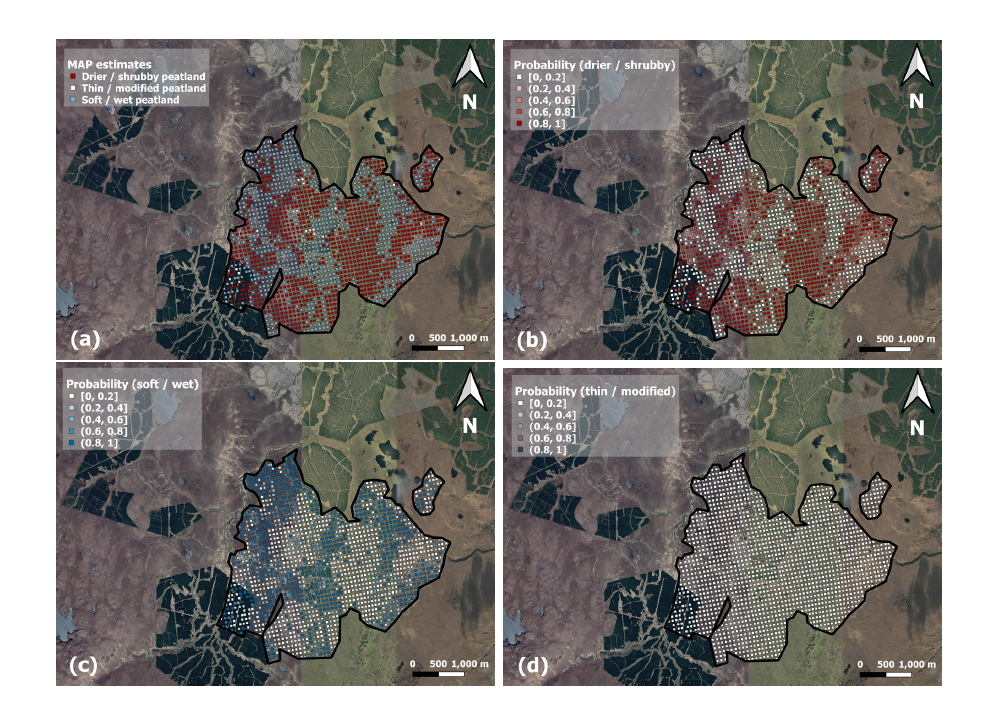}
\end{center}
		\caption{(a) Maximum a posteriori (MAP) estimate for peatland condition in Cross Lochs, (b) probability of drier/shrubby peatland, (c) probability of soft/wet peatland, (d) probability of thin/modified peatland. Base map data: ©2022 Google. Base map imagery: ©2022 CNES/Airbus, Getmapping plc, Landsat/Copernicus, Maxar Technologies.}
\end{figure}

\begin{figure}[t]
\begin{center}
\includegraphics[width=16cm]{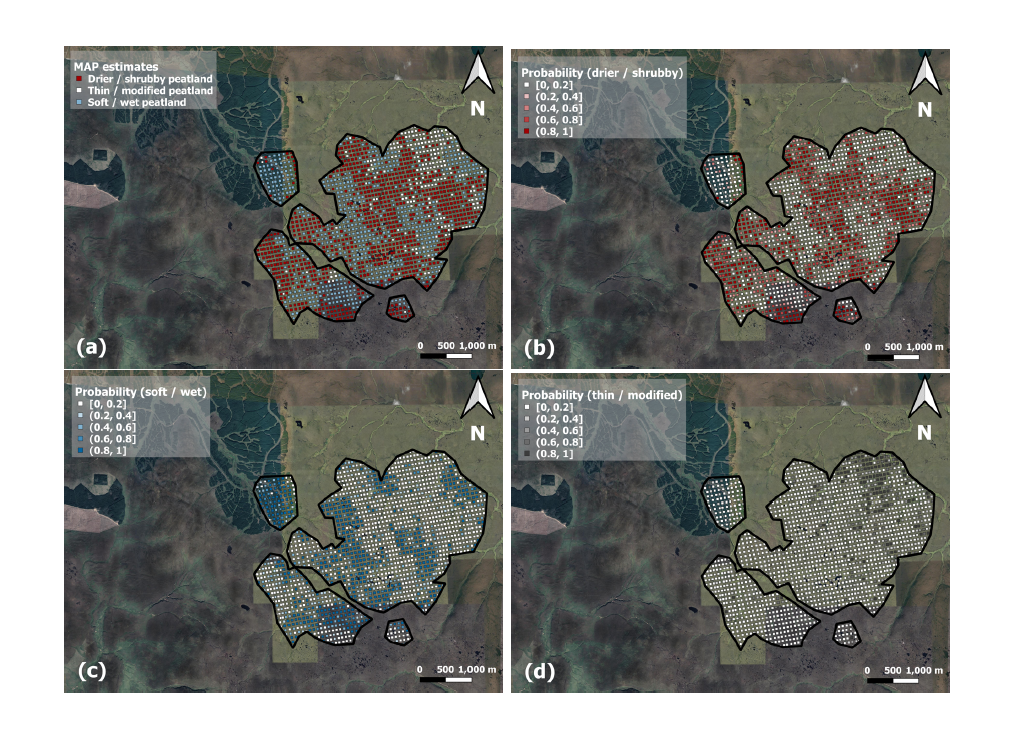}
\end{center}

		\caption{(a) Maximum a posteriori (MAP) estimate for peatland condition in Knockfin, (b) probability of drier/shrubby peatland, (c) probability of soft/wet peatland, (d) probability of thin/modified peatland. Base map data: ©2022 Google. Base map imagery: ©2022 CNES/Airbus, Getmapping plc, Landsat/Copernicus, Maxar Technologies.}
\end{figure}

\begin{figure}[t]
\begin{center}
\includegraphics[width=16cm]{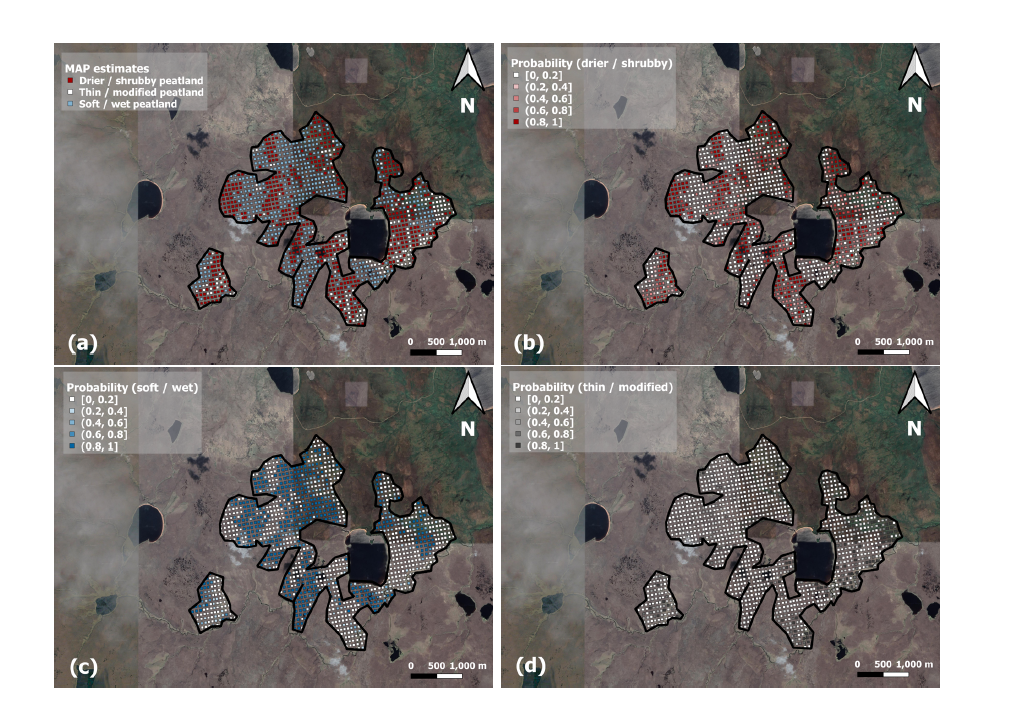}
\end{center}
		\caption{(a) Maximum a posteriori (MAP) estimate for peatland condition in Loch Calium, (b) probability of drier/shrubby peatland, (c) probability of soft/wet peatland, (d) probability of thin/modified peatland. Base map data: ©2022 Google. Base map imagery: ©2022 CNES/Airbus, Getmapping plc, Landsat/Copernicus, Maxar Technologies.}
\end{figure}

\begin{figure}[t]
\begin{center}
\includegraphics[width=16cm]{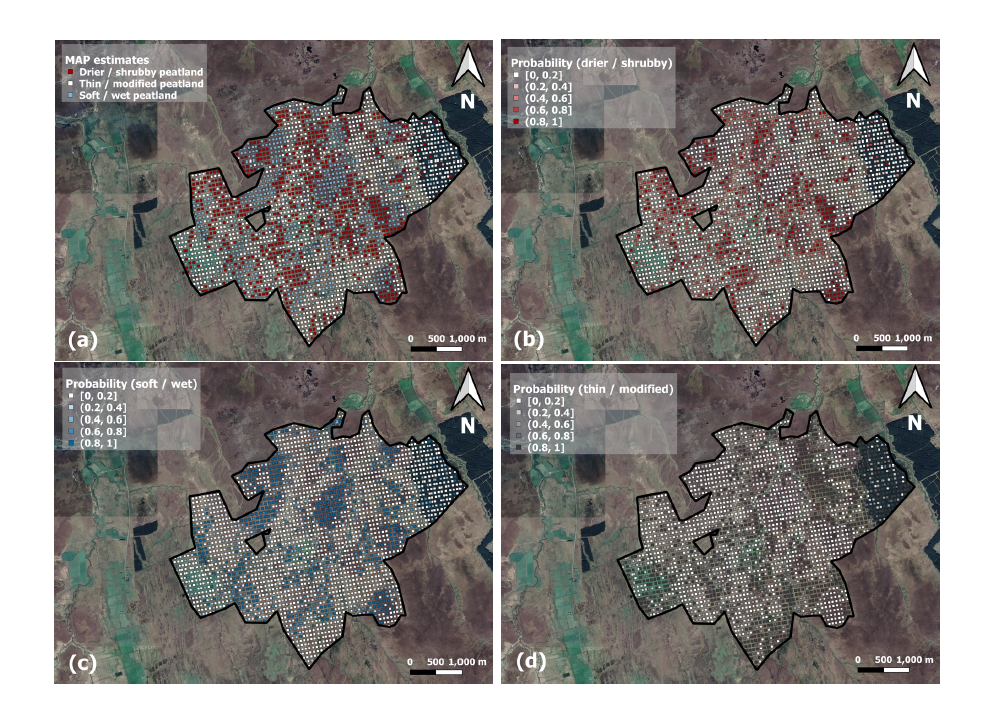}
\end{center}
		\caption{(a) Maximum a posteriori (MAP) estimate for peatland condition in Munsary, (b) probability of drier/shrubby peatland, (c) probability of soft/wet peatland, (d) probability of thin/modified peatland. Base map data: ©2022 Google. Base map imagery: ©2022 CNES/Airbus, Getmapping plc, Landsat/Copernicus, Maxar Technologies.}
\end{figure}

\end{document}